# The Effect of Non-Local Electrical Conductivity on Near-Field Radiative Heat Transfer between Graphene Sheets

Saman Zare[*,**], Behrad Zeinali Tajani[*], and Sheila Edalatpour[*,**]

[*]Department of Mechanical Engineering, University of Maine, Orono, Maine 04469, USA

[**]Frontier Institute for Research in Sensor Technologies, University of Maine, Orono, Maine 04469, USA

**Abstract**

Graphene's near-field radiative heat transfer is determined from its electrical conductivity, which is commonly modeled using the local (wavevector independent) Kubo and Drude formulas. In this letter, we analyze the non-locality of the graphene electrical conductivity using the Lindhard model combined with the Mermin relaxation time approximation. We also study how the variation of the electrical conductivity with the wavevector affects near-field radiative conductance between two graphene sheets separated by a vacuum gap. It is shown that the variation of the electrical conductivity with the wavevector, $k_\rho$, is appreciable for $k_\rho$s greater than $100k_0$, where $k_0$ is the magnitude of the wavevector in the free space. The Kubo model is obtained by assuming $k_\rho \to 0$, and thus is not valid for $k_\rho > 100k_0$. The Kubo electrical conductivity provides an accurate estimation of the spectral radiative conductance between two graphene sheets except for around the surface-plasmon-polariton frequency of graphene and at separation gaps smaller than 20 nm where there is a non-negligible contribution from electromagnetic modes with $k_\rho > 100k_0$ to the radiative conductance. The Drude formula proves to be inaccurate for modeling the electrical conductivity and radiative conductance of graphene except for at temperatures much below the Fermi temperature and frequencies much smaller than $\frac{2\mu_c}{\hbar}$, where $\mu_c$ and $\hbar$ are the chemical potential and reduced Planck's constant, respectively. It is also shown that the electronic scattering processes should be considered in the Lindhard model properly, such that the local electron number is conserved. A simple substitution of $\omega$ by $\omega + i\gamma$ ($\omega$, $i$, and $\gamma$ being the angular frequency, imaginary unit, and scattering rate, respectively) in the collision-less Lindhard model does not satisfy the conservation of the local electron number and results in significant errors in computing the electrical conductivity and radiative conductance of graphene.

# I. INTRODUCTION

Radiative heat transfer is in the near-field regime when the separation distance of the heat exchanging media is comparable to or less than the wavelength of thermal radiation. Otherwise, radiative heat transfer is said to be in the far-field regime. Near-field radiative heat transfer (NFRHT) exceeds the far-field blackbody limit by several orders of magnitude due to an extraneous contribution from evanescent waves that are confined in a distance approximately equal to the thermal wavelength from the emitter. When the heat exchanging media support surface modes such as surface phonon polaritons and surface plasmon polaritons (SPPs), NFRHT can become quasi-monochromatic. Near-field radiative heat transfer is very promising for applications such as thermophotovoltaic conversion of thermal to electrical energy [1,2], thermal rectification [3], and near-field photonic cooling [4] to name only a few. Most of the near-field applications require heat exchanging media that support surface modes in the infrared region of the electromagnetic spectrum, where the surface modes can be thermally excited. The ability to tune the spectral location of the surface modes is also highly desired. Graphene is a great candidate for NFRHT applications as it supports SPPs in the infrared region, and the spectral location of the SPP modes can significantly be tuned by changing the chemical potential of graphene via applying a bias voltage. So far, graphene has been proposed for heat transfer enhancement [5-34], active control and switching of radiative heat transfer [18,21,27,29,31,35-42], thermophotovoltaic power generation [37,43,44], heat transfer suppression [8,17,33], ultrafast modulation of heat transfer [45], and active control of the direction of heat flow [46].

Most of the theoretical studies on graphene's NFRHT are based on using the local Kubo [5-18,20,22,24-29,32-38,40-45] and Drude [19,23,30] models for the electrical conductivity of graphene. These two local models are obtained by making several simplifying assumptions as will be discussed in Section III. For example, when deriving the Kubo and Drude electrical conductivities, it is assumed that the parallel component of the wavevector, $k_\rho$, approaches zero [47,48]. However, NFRHT is mediated by electromagnetic waves with various $k_\rho$s ranging from 0 to infinity. Particularly, the contribution from waves with large $k_\rho$s can be significant when the SPP and hyperbolic modes are excited. It is not clear whether the Kubo and Drude electrical conductivities of graphene can accurately estimate NFRHT in systems involving this material. In this paper, we study the variation of graphene's electrical conductivity with $k_\rho$ using the non-local Lindhard-Mermin model and investigate how this variation can affect NFRHT between two graphene sheets separated by a vacuum gap. It is shown that the local Kubo formula can provide an acceptable estimation of the electrical conductivity of graphene for $k_\rho < 100k_0$, where $k_0$ is the magnitude of the wavevector in the free space. Inversely, the Drude model is valid only at temperatures much smaller than the Fermi temperature and frequencies much smaller than $\frac{2\mu_c}{\hbar}$, where $\mu_c$ and $\hbar$ are the chemical potential and reduced Planck's constant, respectively. It is also shown that except for around the SPP resonance frequency of graphene and at distances smaller than ~20 nm, where there is a non-negligible contribution from modes with $k_\rho > 100k_0$ to NFRHT, the radiative conductance between two graphene sheets obtained using the local Kubo formula agrees well with the one found from the non-local Lindhard-Mermin model.

## II. PROBLEM DEFINITION

The problem under consideration is schematically shown in Fig. 1. Two graphene sheets with chemical potential $\mu_c$ are separated by a vacuum gap of size $d$. One of the graphene sheets is at temperature $T$, while the other is kept at $T + \delta T$, where $\delta T$ is an infinitesimal temperature difference. The near-field radiative conductance of the graphene sheets is desired in this study.

The total (integrated over frequency) radiative conductance, $G_{tot}$, is defined as $G_{tot} = \lim_{\delta T \to 0} \frac{q_{tot}}{\delta T}$, where $q_{tot}$ is the total radiative heat flux between the two graphene sheets. The total radiative conductance is obtained by integrating the spectral (frequency dependent) radiative conductance, $G_\omega$, over frequency as:

$$G_{tot} = \int_0^\infty G_\omega d\omega \qquad (1)$$

where $\omega$ is the angular frequency. The spectral radiative conductance is found by integrating the spectral radiative conductance per unit parallel component of the wavevector, $G_{\omega,k_\rho}$, over the parallel component of the wavevector, i.e.,

$$G_\omega = \int_0^\infty G_{\omega,k_\rho} dk_\rho \qquad (2)$$

The term $G_{\omega,k_\rho}$ in Eq. 2 is found using the framework of fluctuational electrodynamics as [49]:

$$G_{\omega,k_\rho} = \begin{cases} \frac{k_\rho}{4\pi^2} \frac{\partial \Theta(\omega,T)}{\partial T} \sum_{\zeta=\text{TE,TM}} \frac{\left(1-|r^\zeta|^2-|t^\zeta|^2\right)^2}{\left|1-(r^\zeta)^2 e^{2ik_{0z}d}\right|^2}, & k_\rho < k_0 \\ \frac{k_\rho}{\pi^2} \frac{\partial \Theta(\omega,T)}{\partial T} \sum_{\zeta=\text{TE,TM}} \frac{(\text{Im}[r^\zeta])^2 e^{-2k_{0z}''d}}{\left|1-(r^\zeta)^2 e^{2ik_{0z}d}\right|^2}, & k_\rho > k_0 \end{cases} \qquad (3)$$

where $\Theta(\omega,T) = \frac{\hbar\omega}{e^{\hbar\omega/k_B T}-1}$ is the mean energy of an electromagnetic state ($k_B$ is the Boltzmann constants), $\zeta$ refers to the polarization state which can be transverse electric (TE) or transverse magnetic (TM), $r^\zeta$ and $t^\zeta$ are the reflection and transmission coefficients at the interface of graphene and vacuum for polarization $\zeta$, respectively, and $k_{0z} = k_{0z}' + ik_{0z}''$ is the z-component (as shown in Fig. 1, the z-axis is normal to the graphene sheets) of the wavevector in the vacuum found as $k_{0z} = \sqrt{k_0^2 - k_\rho^2}$. It should be noted that the variation of $r^\zeta$ and $t^\zeta$ with temperature is neglected when taking the derivative of the heat flux with respect to temperature in Eq. 3, which is appropriate for NFRHT applications except for thermal rectification [50]. The reflection and transmission coefficients at the graphene-vacuum interface can be found for the TE and TM polarizations as [51]:

$$r^{TE} = \frac{-\mu_0 \sigma \omega}{2k_{0z} + \mu_0 \sigma \omega} \tag{4a}$$

$$r^{TM} = \frac{\sigma k_{0z}}{2\varepsilon_0 \omega + \sigma k_{0z}} \tag{4b}$$

$$t^{TE} = \frac{2k_{0z}}{2k_{0z} + \mu_0 \sigma \omega} \tag{4c}$$

$$t^{TM} = \frac{2\varepsilon_0 \omega}{2\varepsilon_0 \omega + \sigma k_{0z}} \tag{4d}$$

where $\varepsilon_0$ and $\mu_0$ are the permittivity and permeability of the vacuum, respectively, and $\sigma$ is the electrical conductivity of graphene. Equations 4a-4d are obtained by assuming an isotropic electrical conductivity for graphene, which is valid except for electromagnetic modes with $k_\rho \gg 300k_0$ [52] or in the presence of a magnetic field or a drift current [31,39].

### III. GRAPHENE'S ELECTRICAL CONDUCTIVITY MODELS

The electrical conductivity of graphene is the key parameter for determining its NFRHT. Three electrical conductivity models, namely, Drude, Kubo, and Lindhard models, have been used for studying NFRHT in graphene-based materials. In this section, we briefly discuss these three models. An extensive review of various electrical conductivity models for graphene can be found in Ref. [53].

### A. Kubo Model

The local Kubo model is obtained using the linear response theory and the Kubo formula under the assumptions that $k_\rho \to 0$, $\omega \gg k_\rho v_F$, and $\omega \gg \gamma$, where $v_F$ is the Fermi velocity and $\gamma$ is a phenomenological parameter called the scattering rate accounting for the electronic scattering processes [47,48,54]. The Kubo electrical conductivity of graphene is written as the summation of a contribution from the intraband transitions of electrons, $\sigma^{\text{intra}}$, and one from the interband transitions of electrons, $\sigma^{\text{inter}}$, i.e.,

$$\sigma(\omega, T, \mu_c) = \sigma^{\text{intra}}(\omega, T, \mu_c) + \sigma^{\text{inter}}(\omega, T, \mu_c) \tag{5a}$$

where $T$ is the temperature, and

$$\sigma^{\text{intra}}(\omega, T, \mu_c) = \frac{4i\sigma_0}{\pi\hbar(\omega+i\gamma)}\left[\mu_c + 2k_BT\ln\left(1 + e^{-\mu_c/k_BT}\right)\right] \tag{5b}$$

while

$$\sigma^{\text{inter}}(\omega, T, \mu_c) = \sigma_0\left[G(\hbar\omega/2) + i\frac{4\hbar\omega}{\pi}\int_0^\infty \frac{G(E) - G(\hbar\omega/2)}{(\hbar\omega)^2 - 4E^2}dE\right] \tag{5c}$$

In Eqs. 5b and 5c, $i$ is the imaginary unit, and $\sigma_0$ and $G$ are defined as:

$$G(x) = \frac{\sinh\left(\frac{x}{k_BT}\right)}{\cosh\left(\frac{\mu_c}{k_BT}\right) + \cosh\left(\frac{x}{k_BT}\right)} \tag{5d}$$

$$\sigma_0 = e^2/(4\hbar) \tag{5e}$$

where $e$ is the electron charge. The scattering rate is the inverse of the relaxation time, i.e., $\gamma = \tau^{-1}$, where $\tau$ is the relaxation time. The relaxation time linearly varies with the chemical potential as [55]:

$$\tau = \frac{m\mu_c}{ev_F^2} \tag{6}$$

In Eq. 6, $m$ is the carrier mobility with a value of 1000-230000 cm$^2$/Vs depending on the method used for fabricating graphene. In this study, $v_F = 9.5\times10^5$ m/s [51] and $m = 10000$ cm$^2$/Vs [55] are considered for Fermi velocity and carrier mobility of graphene, respectively. The intraband contribution to the Kubo electrical conductivity is dominant when $\hbar\omega \ll 2\mu_c$, while the interband transitions become significant when $\hbar\omega > 2\mu_c$ [53].

### B. Drude Model

At low frequencies (i.e., when $\omega \ll \frac{2\mu_c}{\hbar}$), the interband contribution to the Kubo electrical conductivity can be neglected. If the temperature is low compared to the Fermi temperature (i.e., if $T \ll \frac{\mu_c}{k_B}$), the second term in the intraband electrical conductivity is also negligible. In this case, the Kubo electrical conductivity is simplified to the following equation which is referred to as the Drude model [48]:

$$\sigma(\omega, \mu_c) = \frac{4i\sigma_0 \mu_c}{\pi\hbar(\omega + i\gamma)} \tag{7}$$

### C. Lindhard Model

The Lindhard electrical conductivity of graphene is a non-local model which follows from a quantum mechanical description of the material using the self-consistent linear response theory [48,53,56,57]. The graphene electrical conductivity can be related to its polarizability, $\Pi$, as [48,53]:

$$\sigma(k_\rho, \omega, T, \mu_c) = i\frac{\omega}{k_\rho^2} \Pi(k_\rho, \omega, T, \mu_c) \tag{8}$$

The polarizability of graphene in the collision-less Lindhard model is given by [48,53,57,58]:

$$\Pi(k_\rho, \omega, T, \mu_c) = 4e^2 \lim_{\eta \to 0^+} \sum_{n,n'=\pm 1} \int \frac{d^2\mathbf{q}}{(2\pi)^2} \left(\frac{1+nn'\cos\theta}{2}\right) \frac{f_{n',\mathbf{q}+\mathbf{k}_\rho} - f_{n,\mathbf{q}}}{\epsilon_{n',\mathbf{q}+\mathbf{k}_\rho} - \epsilon_{n,\mathbf{q}} - \hbar(\omega + i\eta)} \tag{9a}$$

where $\eta$ is a small number accounting for the Landau damping [48,57], $\mathbf{k}_\rho$ is the parallel (to the graphene sheet) component of the wavevector, $\mathbf{q}$ denotes a vector in the $\mathbf{k}_\rho$-space, and

$$\epsilon_{n,\mathbf{q}} = n\hbar v_F |\mathbf{q}| \tag{9b}$$

$$f_{n,\mathbf{q}} = \left(1 + \exp(\epsilon_{n,\mathbf{q}} - \mu_c)/k_B T\right)^{-1} \tag{9c}$$

$$\cos\theta = \frac{\mathbf{q} \cdot (\mathbf{q} + \mathbf{k}_\rho)}{|\mathbf{q}||\mathbf{q} + \mathbf{k}_\rho|} \tag{9d}$$

As indicated by Eq. 9d, $\theta$ is the angle between vectors $\mathbf{q}$ and $\mathbf{q} + \mathbf{k}_\rho$. It is shown that the Lindhard model can reproduce the polarizability of graphene found from ab initio calculations for frequencies below $4.5\times10^{15}$ rad/s (or wavelengths above 0.42 μm) [53], the spectrum region in which thermal radiation is typically located. It should be noted that the electronic scattering processes are ignored in the collision-less Lindhard model [48,53,57]. The electronic collisions can be accounted for in the collision-less Lindhard model using the relaxation-time approximation (RTA) via the scattering rate, $\gamma$ [48,59]. Two approaches have been used for including the electronic collisions in the Lindhard model. In the first approach, the angular frequency $\omega$ in the collision-less Lindhard formula (Eq. 9a) is simply replaced by $\omega + i\gamma$ [53,59]. This approach is known to violate the conservation of the local electron number [48,59,60] and is referred to as the Lindhard model hereafter. In the second approach, which was proposed by Mermin, the collisions relax the electronic density to an equilibrium distribution with a shifted chemical potential such that the local electron number is conserved [48,60]. Using this approach, which is referred to as the Lindhard-Mermin model in this paper, the polarizability of graphene is modified as [48,60]:

$$\Pi_\gamma(k_\rho,\omega,T,\mu_c,\gamma) = \frac{(\omega+i\gamma)\Pi(k_\rho,\omega+i\gamma,T,\mu_c)}{\omega+i\gamma[\Pi(k_\rho,\omega+i\gamma,T,\mu_c)/\Pi(k_\rho,0,T,\mu_c)]} \quad (10)$$

where $\Pi(k_\rho,\omega+i\gamma,T,\mu_c)$ is given by Eq. 9a. A comparison of the discussed electrical conductivity models is presented in Section IV.

## IV. RESULTS AND DISCUSSION

The real part and the absolute value of the imaginary part of graphene's electrical conductivity as predicted using the Lindhard-Mermin (Eqs. 9 and 10), Lindhard (Eq. 9), Drude (Eq. 7), and Kubo (Eq. 5) models are shown in Fig. 2 for three cases. In Case 1, $\mu_c = 0.05$ eV and $T = 300$ K (Figs. 2a and 2b). In Case 2, $\mu_c$ is increased to 0.1 eV, while $T$ is kept at 300 K (Figs. 2c and 2d). Case 3 is concerned with a $\mu_c$ of 0.1 eV and an enhanced $T$ of 1000 K (Figs. 2e and 2f). The Lindhard-Mermin and Lindhard electrical conductivities are computed at a small $k_\rho$ of $0.05k_0$, while the Kubo and Drude models do not account for the variation of the electrical conductivity with $k_\rho$. It is seen from Fig. 2 that the real part of the electrical conductivity, Re[$\sigma$], found from the Kubo model agrees with the one obtained using the Lindhard-Mermin model for small $k_\rho$s. The maximum difference between Re[$\sigma$] found from these two models is only 9.1% (the difference depends on the frequency) in Fig. 2a, 10.9% in Fig. 2c, and 3.8% in Fig. 2e. The agreement between the Kubo and the Lindhard-Mermin models can be explained by considering the assumptions made when deriving the Kubo electrical conductivity as discussed in Section III.A. In the Kubo electrical conductivity, it is assumed that $k_\rho \to 0$, $\omega \gg k_\rho v_F$, and $\omega \gg \gamma$. Considering that $v_F \approx \frac{c_0}{300}$ ($c_0$ being the speed of light in vacuum) [51], the second assumption is valid when $k_\rho \ll 300 k_0$, while the third assumption holds true for $\omega \gg 10^{13}$ rad/s. Since these two conditions are satisfied for the cases presented in Fig. 2, the Kubo formula predicts Re[$\sigma$] with an acceptable accuracy in this figure. It is also seen from Fig. 2 that Re[$\sigma$] as predicted using the Lindhard model agrees with the one found from the Lindhard-Mermin model only at large frequencies. The reason is that electron scattering by impurities and lattice defects, which has not been appropriately accounted for in the Lindhard model, is significant mostly at low frequencies [61]. The large difference between the Lindhard and Lindhard-Mermin models at low frequencies highlights the importance of proper consideration of electronic scattering processes in the collision-less Lindhard formula. Figure 2 also shows that except for at low frequencies and temperatures, the Drude model cannot accurately estimate the electrical conductivity of graphene. As mentioned in Section III.A, the Drude model is valid under the assumptions that $\omega \ll \frac{2\mu_c}{\hbar}$ and $T \ll \frac{\mu_c}{k_B}$. The former assumption, which ensures no contribution from the interband transitions to the electrical conductivity, is valid only for $\omega \ll 1.5\times10^{14}$ rad/s when $\mu_c = 0.05$ eV (Figs. 2a and 2b) and for $\omega \ll 3.0\times10^{14}$ rad/s when $\mu_c = 0.1$ eV (Figs. 2c-2f). As such, the Drude model fails at frequencies comparable to or greater than $10^{14}$ rad/s. It should be pointed out that the frequency below which the Drude model is valid is proportional to $\mu_c$ ($\omega \ll \frac{2\mu_c}{\hbar}$) and thus increases as $\mu_c$ increases (e.g., compare Figs. 2a and 2c). Assuming $T \ll \frac{\mu_c}{k_B}$ in the derivation of the Drude model ensures that the temperature dependence of the electrical conductivity is negligible. When $\mu_c = 0.1$ eV, $T \ll 1160$ K is required for satisfying this condition. Since this condition is not met for Case 3 with $T = 1000$ K, the Drude model deviates from the Lindhard-Mermin model for this case at all frequencies (Figs. 2e and 2f). The results presented in Fig. 2 contradicts the conclusion made in Ref. [53] that the Drude model accurately predicts the electrical conductivity of graphene for $k_\rho < 300k_0$. It is seen from Fig. 2b that the imaginary part of the electrical conductivity, Im[$\sigma$], obtained using the Kubo formula for Case 1 differs from the one found using the Lindhard-Mermin model at large frequencies. However, as the chemical potential and temperature increase, the Kubo predictions for Im[$\sigma$] converge to the ones obtained from the Lindhard-Mermin model (Figs. 2c-2f).

The effect of $k_\rho$ on the electrical conductivity of graphene is studied in Fig. 3. In this figure, Re[$\sigma$] and Im[$\sigma$] predicted using the Lindhard-Mermin model for various $k_\rho$s from $0.05k_0$ to $200k_0$ are shown for Cases 1 to 3. The Kubo electrical conductivity, which is $k_\rho$-independent, is also shown in Fig. 3 for comparison. When $k_\rho \ll 100k_0$, the variation of Re[$\sigma$] with $k_\rho$ is modest and the Kubo formula can be used for predicting the electrical conductivity with acceptable accuracy. For example, Re[$\sigma$] in Fig. 3a changes only between 0.03% and 13.25% (depending on the frequency) when $k_\rho$ increases from $0.05k_0$ to $100k_0$. The variation of Re[$\sigma$] with $k_\rho$ becomes significantly stronger as $k_\rho$ increases, such that the Kubo formula cannot be used for $k_\rho \geq 100k_0$. When $k_\rho$ increases from $100k_0$ to $200k_0$, Re[$\sigma$] in Fig. 3a changes by up to 54%. The same conclusion can be made using Figs. 2c-2f which show the variation of $\sigma$ with $k_\rho$ for Case 2 with a higher chemical potential of $\mu_c = 0.1$ eV (Figs. 3c and 3d) and Case 3 with a higher temperature of $T = 1000$ K (Figs. 3e and 3f).

To study the effect of the variation of $\sigma$ with $k_\rho$ on NFRHT, the spectral radiative conductance between two graphene sheets is calculated using the discussed electrical conductivity models. The radiative conductance at $d = 50$ nm is shown for Cases 1 to 3 in Figs. 4a to 4c, respectively, while Fig. 4d shows the radiative conductance for Case 3 at an increased gap size of $d = 500$ nm. It is seen from Fig. 4 that the radiative conductance has a peak in the considered frequency region which is due to thermal emission of SPPs. The SPP peak has a major contribution to the total radiative conductance. Figure 4 shows a great agreement between the Kubo and Lindhard-Mermin radiative conductances except for around the SPP frequency. At the SPP frequency, the Kubo formula overestimates the magnitude of the radiative conductance by 17% for Case 1 shown in Fig. 4a. This difference reduces to 11.7% when the chemical potential is increased to 0.1 eV in Fig. 4b, and to 6.5% when the temperature increases to 1000 K in Fig. 4c. The reason for the difference between the Kubo and Lindhard-Mermin radiative conductances around the SPP frequency can be explained by considering the distribution of the spectral radiative conductance over $k_\rho$. The spectral radiative conductance per unit wavevector, $G_{\omega,k_\rho}$, is plotted versus $\omega$ and $k_\rho/k_0$ for the Lindhard-Mermin and Kubo models in Fig. 5. The dispersion relation of the graphene SPPs is also shown in Fig. 5. The dispersion relation of the SPPs for two graphene sheets separated by a gap of size $d$ (Fig. 1) is split into two branches: One branch corresponding to the optical mode given by $\frac{1}{\kappa}\tanh\left(\frac{\kappa d}{2}\right) + \frac{1}{\kappa} + \frac{i\sigma}{\omega\varepsilon_0} = 0$, and one corresponding to the acoustic mode found from $\frac{1}{\kappa}\coth\left(\frac{\kappa d}{2}\right) + \frac{1}{\kappa} + \frac{i\sigma}{\omega\varepsilon_0} = 0$, where $\kappa = \sqrt{k_\rho^2 - k_0^2}$ [48,62]. Figure 5a shows $G_{\omega,k_\rho}$ for Case 1 at $d = 50$ nm (corresponding to $G_\omega$ in Fig. 4a). As it is seen from this figure, there is a contribution from electromagnetic waves with $k_\rho > 100k_0$ to the radiative conductance around the SPP frequency. Since the difference between the Kubo and Lindhard-Mermin electrical conductivities is appreciable for $k_\rho > 100k_0$, the spectral radiative conductances obtained using these two models do not agree around the SPP frequency. The contribution of the electromagnetic waves with $k_\rho > 100k_0$ to the radiative conductance reduces as $\mu_c$, $T$ and $d$ increase (see Figs. 5b to 5d), and so does the difference between the Kubo and Lindhard-Mermin radiative conductances around the SPP frequency. It is also seen from Fig. 5 that the dispersion relation of graphene's SPPs obtained using the Kubo model deviates significantly from the one found from the Lindhard-Mermin electrical conductivity at large wavevectors for which the Kubo formula is not valid. Figure 4 also shows that the Lindhard and Drude electrical conductivities cannot accurately predict the spectral radiative conductance and thus are not recommended for modeling NFRHT for graphene-based materials.

The total radiative conductance, $G_{tot}$, for the two graphene sheets as predicted using the discussed electrical conductivity models is presented in Fig. 6. Figure 6a shows the total radiative conductance versus $\mu_c$ for $T = 300$ K and $d = 50$ nm. The total radiative conductances obtained using the Kubo and Lindhard-Mermin models agree well for all considered chemical potentials. The Kubo total radiative conductance is within ~ 5% of the one found from the Lindhard-Mermin model. Inversely, the total radiative conductances predicted using the Drude and Lindhard models are significantly different from the one found from the Lindhard-Mermin electrical conductivity. It is seen from Fig. 6a that the Lindhard-Mermin total radiative conductive non-monotonically varies by ~40 times as the chemical potential changes in the range of 0.05 eV to 0.5 eV, which is very promising for active control of radiative heat transfer. The total radiative conductance versus $T$ is plotted in Fig. 6b for $\mu_c = 0.05$ eV and $d = 50$ nm. Figure 6b also demonstrates that the Kubo model provides an acceptable estimation for the total radiative conductance. In this figure, the difference between the Kubo and Lindhard-Mermin radiative conductances at $T = 300$ K is 5.5%. This difference constantly decreases with increasing temperature such that the two models are different by only 1.3% at $T = 1000$ K. Figure 6b also shows that the total radiative conductance increases almost linearly with temperature by ~3 times as the temperature increases from 300 K to 1000 K. The effect of $d$ on the accuracy of the discussed electrical conductivity models for predicting the total radiative conductance is shown in Fig. 6c, where the total radiative conductance is plotted versus $d$ for $\mu_c = 0.05$ eV and $T = 300$ K. This figure shows that the difference between the total radiative conductances predicted using the Kubo and Lindhard-Mermin models increases considerably as the separation gap decreases. For example, the difference is 2.5% for $d = 1$ mm, while it increases to 36% at $d = 10$ nm. The reason is that the contribution of thermally emitted waves with $k_\rho > 100 k_0$ to the radiative conductance increases when $d$ decreases (see Figs. 2e-2h). As discussed before, the Kubo electrical conductivity cannot be used when $k_\rho > 100 k_0$. Additionally, Fig. 6c shows that the Drude model cannot also provide an accurate estimation for the far-field radiative conductance.

## V. CONCLUSION

In summary, the Kubo formula can be used for modeling NFRHT with an acceptable accuracy except for at separation gaps smaller than ~20 nm and around the SPP resonance frequency. In this study, the difference between the total radiative conductance found using the Kubo formula and the one obtained using the Lindhard-Mermin electrical conductivity exceeded 20% at gaps smaller than 20 nm. The peak spectral radiative conductance at the SPP resonance predicted using the Kubo formula was different from that of the Lindhard-Mermin model by up to 17%. At small separation gaps or around the SPP frequency of graphene, there is an appreciable contribution from electromagnetic modes with $k_\rho$ greater than $100k_0$ for which the Kubo formula ceases to be valid. As the chemical potential, temperature, and separation distance of the graphene sheets increases, the contribution of large-$k_\rho$ modes to the NFRHT decreases and thus the accuracy of the Kubo formula for predicting the radiative conductance increases. It was also shown that, in contrast to previous findings [53], the Drude model is not valid for modeling NFRHT problems except for $\omega \ll \frac{2\mu_c}{\hbar}$ and $T \ll \frac{\mu_c}{k_B}$. Finally, simple substitution of $\omega$ with $\omega + i\gamma$ in the collision-less Lindhard model cannot accurately capture the effect of electronic scattering processes on graphene's electrical conductivity, as the local electron number is not conserved in this approach. The Mermin relaxation time approximation is recommended for modeling the scattering processes in the graphene's electrical conductivity.

**Acknowledgment**

This work is supported by the National Science Foundation under Grant No. CBET-2046630.

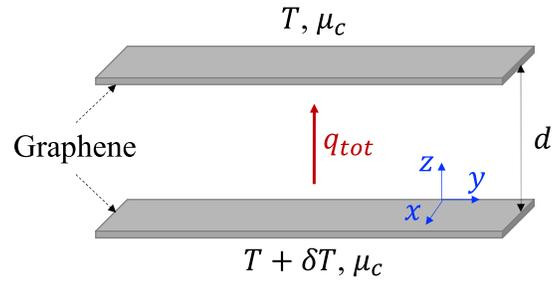

Figure 1 – A schematic of the problem under consideration. Two graphene sheets with a chemical potential of $\mu_c$ are separated by a vacuum gap of size $d$. One of the sheets is at temperature $T$, while the other is at a temperature infinitesimally greater than $T$. The radiative conductance of the graphene sheets is desired.

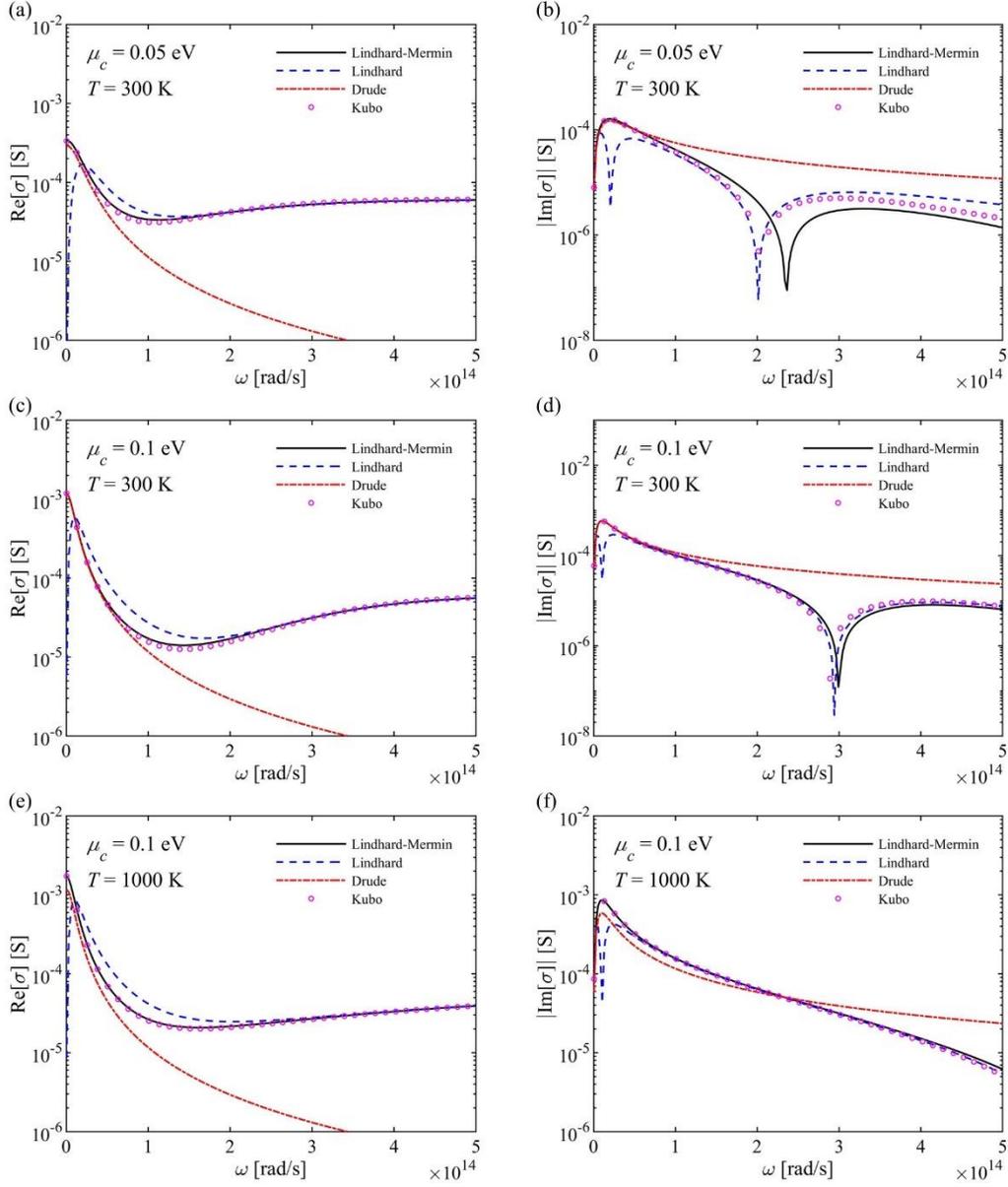

Figure 2 – The real part and the absolute value of the imaginary part of the electrical conductivity of graphene, $\sigma$, as calculated using the Lindhard-Mermin, Lindhard, Drude, and Kubo models. The Lindhard-Mermin and Lindhard electrical conductivities are calculated at $k_\rho = 0.05 k_0$, while the Drude and Kubo models are independent of $k_\rho$. Panels (a) and (b), respectively, show the real and imaginary parts of $\sigma$ for Case 1 with $\mu_c = 0.05$ eV and $T = 300$ K. The same are shown for Case 2 with $\mu_c = 0.1$ eV and $T = 300$ K in Panels (c) and (d), and for Case 3 with $\mu_c = 0.1$ eV and $T = 1000$ K in Panels (e) and (f).

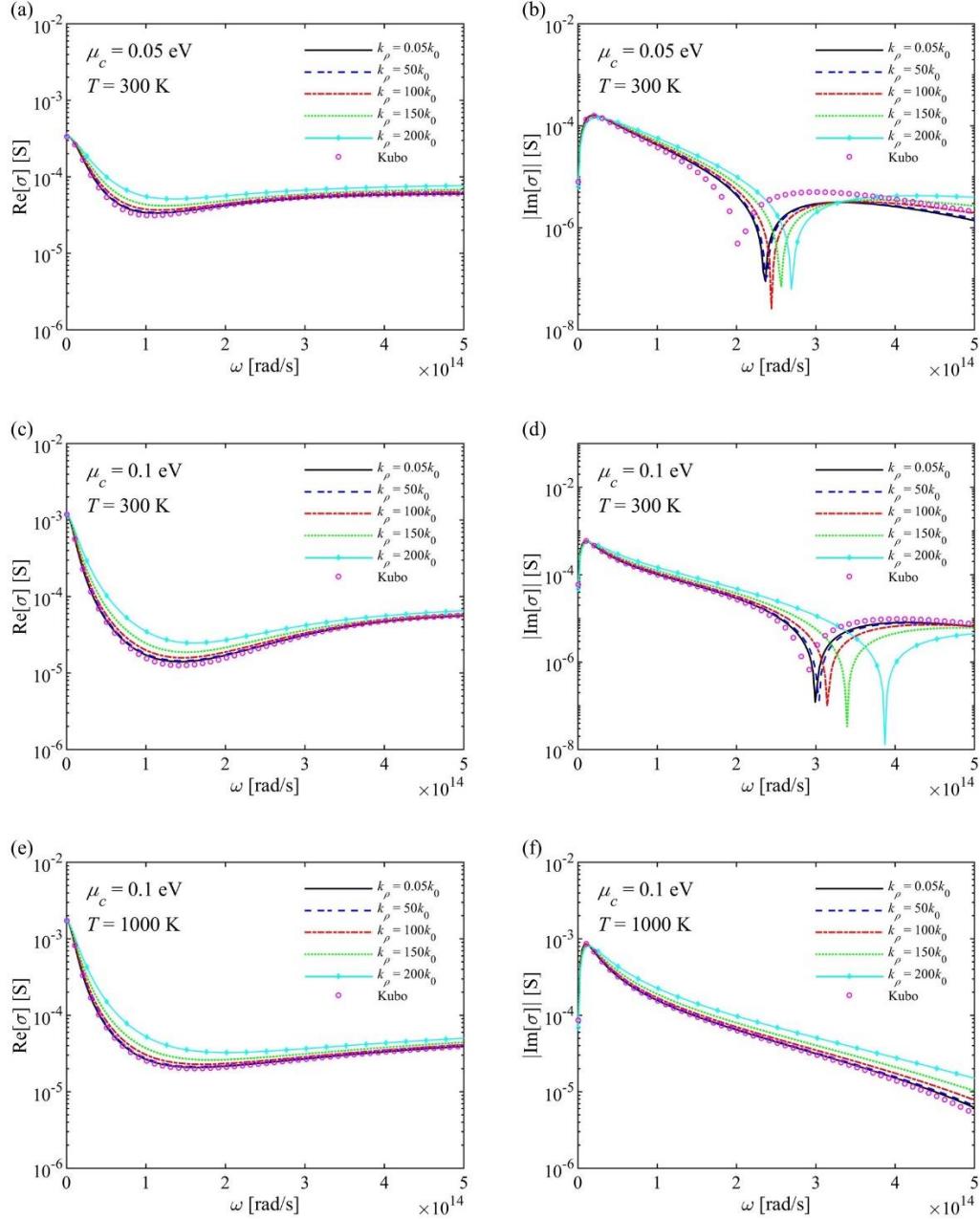

Figure 3 – The real part and the absolute value of the imaginary part of the Lindhard-Mermin and Kubo electrical conductivities of graphene, $\sigma$. The Lindhard-Mermin electrical conductivity is plotted for various $k_\rho$s, while the Kubo model is independent of $k_\rho$. Panels (a) and (b), respectively, show the real and imaginary parts of $\sigma$ for Case 1 with $\mu_c = 0.05$ eV and $T = 300$ K. The same are shown for Case 2 with $\mu_c = 0.1$ eV and $T = 300$ K in Panels (c) and (d), and for Case 3 with $\mu_c = 0.1$ eV and $T = 1000$ K in Panels (e) and (f).

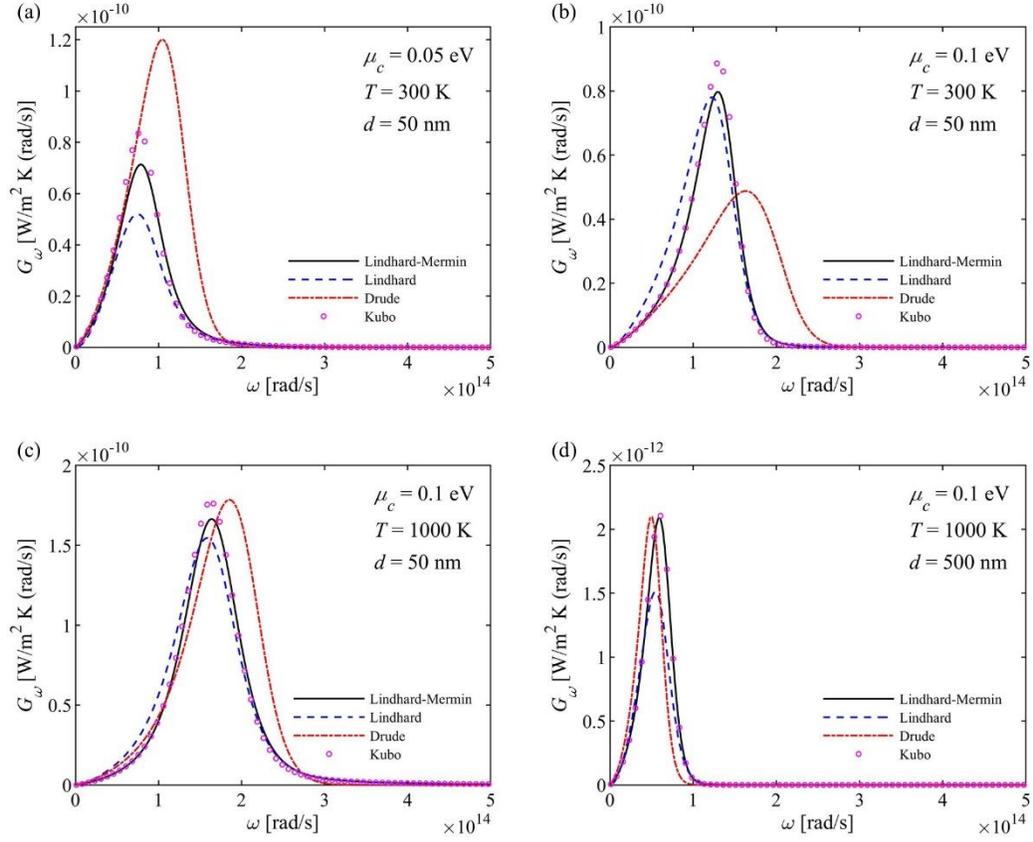

Figure 4 – Spectral (frequency dependent) radiative conductance between two graphene sheets with chemical potential $\mu_c$, temperature $T$, and separation gap $d$ as predicted using the Lindhard-Mermin, Lindhard, Drude, and Kubo electrical conductivities.

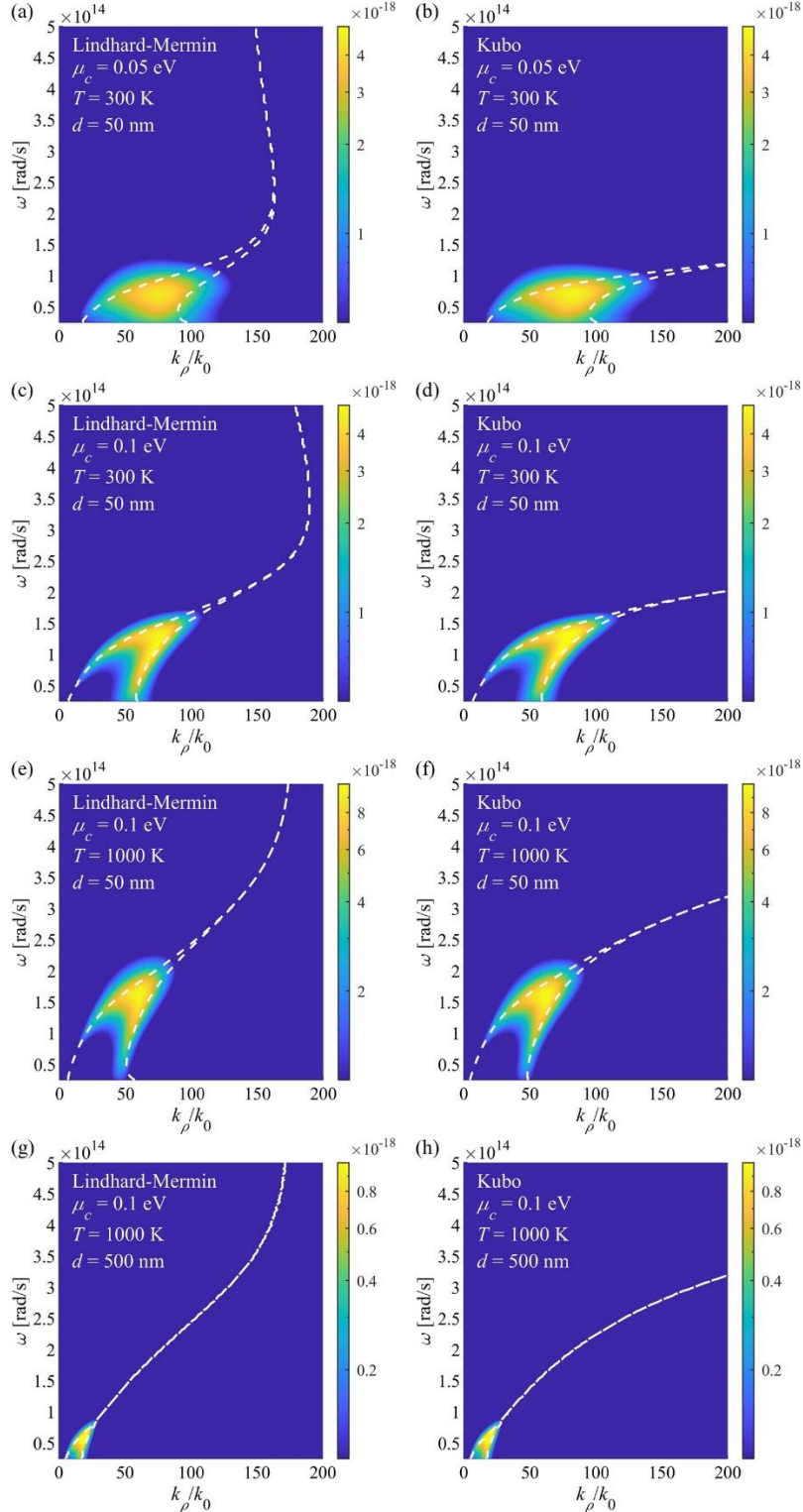

Figure 5 – Spectral radiative conductance per unit $k_\rho$, $G_{\omega,k_\rho}$, for two graphene sheets with a chemical potential of $\mu_c$ and a temperature of $T$ separated by a gap of size $d$ as computed using

the Lindhard-Mermin (Panels (a), (c), (e), and (g)) and Kubo (Panels (b), (d), (f), and (h)) electrical conductivities. The unit for $G_{\omega,k_\rho}$ shown in the color bar is Wm$^{-2}$ (rad/s)$^{-1}$(rad/m)$^{-1}$.

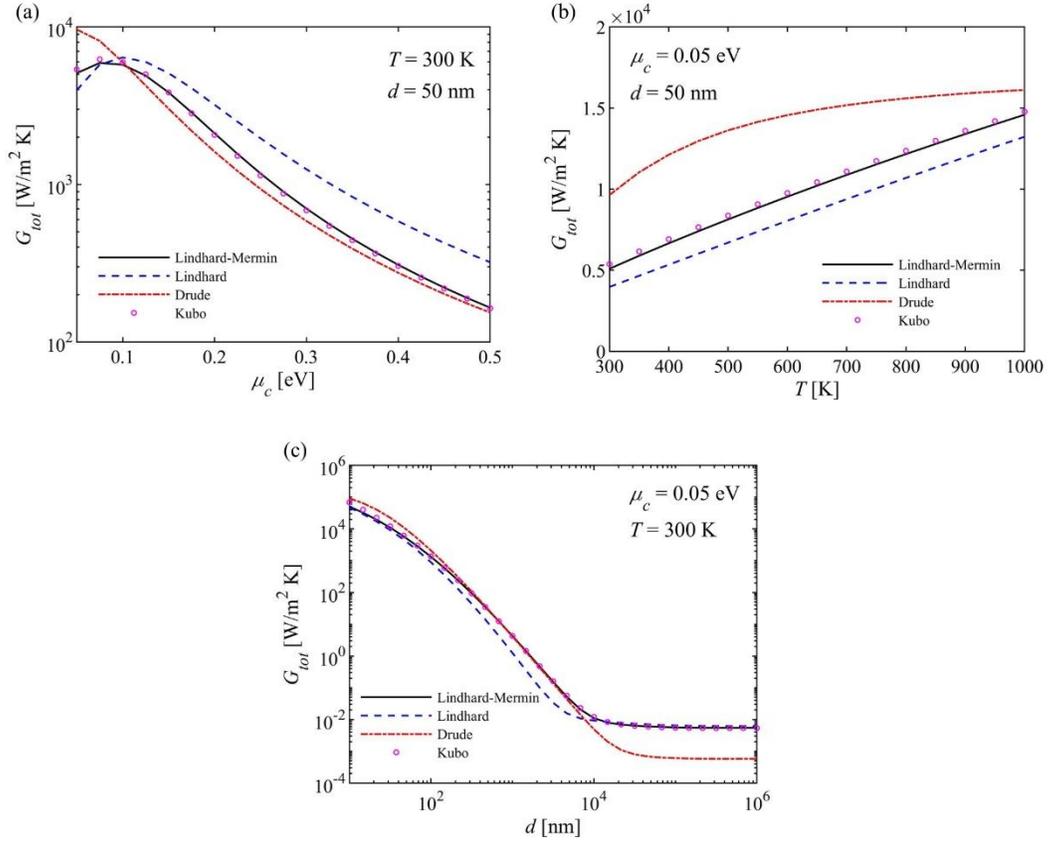

Figure 6 – Total radiative conductance, $G_{tot}$, between two graphene sheets separated by a vacuum gap as predicted using the Lindhard-Mermin, Lindhard, Drude, and Kubo electrical conductivities versus (a) chemical potential, $\mu_c$, (b) temperature, $T$, and (c) separation gap, $d$.